\documentclass[10pt,conference]{IEEEtran}
\IEEEoverridecommandlockouts

\usepackage{cite}
\usepackage{amsmath,amssymb,amsfonts}
\usepackage{algorithmic}
\usepackage{graphicx}
\usepackage{subcaption}
\usepackage{svg}
\usepackage{textcomp}
\usepackage{booktabs}
\usepackage{xcolor}
\usepackage{multirow}
\def\BibTeX{{\rm B\kern-.05em{\sc i\kern-.025em b}\kern-.08em
    T\kern-.1667em\lower.7ex\hbox{E}\kern-.125emX}}

\usepackage{pbalance}

\makeatletter
\newcommand{\linebreakand}{%
  \end{@IEEEauthorhalign}
  \hfill\mbox{}\par
  \mbox{}\hfill\begin{@IEEEauthorhalign}
}
\makeatother
\newcommand{\revision}[1]{\textcolor{black}{#1}}

\usepackage{hyperref}
\hypersetup{
    colorlinks=true,
    allcolors=blue,
    pdftitle={Enhancing Anomaly-Based Intrusion Detection Systems with Process Mining},
    pdfpagemode=FullScreen,
}

\begin{document}
\title{Enhancing Anomaly-Based Intrusion Detection Systems with Process Mining
\thanks{This work was supported by the DEFEDGE project (E53D23016380001) under the PRIN program.}
}

\author{%
\IEEEauthorblockN{Francesco Vitale}
\IEEEauthorblockA{\textit{DIETI} \\
\textit{University of Naples Federico II}\\
Naples, Italy \\
francesco.vitale@unina.it}
\and
\IEEEauthorblockN{Francesco Grimaldi}
\IEEEauthorblockA{\textit{DIETI} \\
\textit{University of Naples Federico II}\\
Naples, Italy \\
francesco.grimaldi3@unina.it}
\linebreakand
\IEEEauthorblockN{Massimiliano Rak}
\IEEEauthorblockA{\textit{DIETI} \\
\textit{University of Naples Federico II}\\
Naples, Italy \\
massimiliano.rak@unina.it}
\and
\IEEEauthorblockN{Nicola Mazzocca}
\IEEEauthorblockA{\textit{DIETI} \\
\textit{University of Naples Federico II}\\
Naples, Italy \\
nicola.mazzocca@unina.it}}

\maketitle

\begin{abstract}
%\note{XMAX: rivedere abstract e allineare a contributi}
Anomaly-based Intrusion Detection Systems (IDSs) ensure protection against malicious attacks on networked systems. While deep learning-based IDSs achieve effective performance, their limited trustworthiness due to black-box architectures remains a critical constraint. Despite existing explainable techniques offering insight into the alarms raised by IDSs, they lack process-based explanations grounded in packet-level sequencing analysis. In this paper, we propose a method that employs process mining techniques to \textit{enhance} anomaly-based IDSs by providing process-based alarm severity ratings and explanations for alerts. \revision{Our method prioritizes critical alerts and maintains visibility into network behavior, while minimizing disruption by allowing misclassified benign traffic to pass.}
%The method exploits process mining techniques to characterize the packet-level sequencing of network protocols and verify the degree of conformance to false-positive behavior, assigning different levels of severity to anomaly-based IDS alarms. 
We apply the method to the publicly available USB-IDS-TC dataset, which includes anomalous traffic affected by different variants of the Slowloris DoS attack. Results show that our method is able to discriminate between low- to very-high-severity alarms while preserving up to 99.94\% recall and 99.99\% precision, effectively discarding false positives while providing different degrees of severity for the true positives.

%Intrusion Detection Systems (IDSs) ensure protection against malicious attacks on networked systems. Many advanced anomaly-based IDSs have been developed through deep learning techniques, alongside eXplainable Artificial Intelligence (XAI) approaches to unveil their black-box behavior. Despite rich literature demonstrating increased evidence in achieving both effective and trustworthy detection, the existing XAI approaches lack a process-based explanation that can uncover the packet-level sequencing of events that led to alarms. In this paper, we propose a method for process-based explanation and rating of anomaly-based IDS alerts, that allows inspecting the control flow of network packets associated with anomaly-based IDS alarms and assigning different levels of alarm severity. We apply the method to the publicly available USB-IDS-TC dataset, which includes anomalous traffic affected by different variants of the Slowloris DoS attack. Results show that our method is able to discriminate between low to very-high-severity alarms while preserving high detection performance.
\end{abstract}

\begin{IEEEkeywords}
Intrusion detection systems, alarm rating, explainable artificial intelligence (XAI), process mining 
\end{IEEEkeywords}

\section{Introduction}
\label{sec:intro}
%\note{xMAX: allinerare intro e abstract}
Intrusion Detection Systems (IDSs) have attracted significant attention in the literature due to their critical role in monitoring and mitigating malicious activity in networked systems. In this paper, we focus on anomaly-based IDSs, which are able to characterize the normal profile of network traffic and identify any deviations through statistical, knowledge-based, and deep learning approaches~\cite{khraisat2019survey}. The literature outlines that the best-performing approaches are predominantly based on deep learning~\cite{aldweesh2020deep}, which has achieved remarkable detection performance across diverse datasets, surpassing 99\% performance figures~\cite{nandanwar2024dlids}. 

Despite the impressive results achieved by anomaly-based IDSs, the literature has also highlighted a fundamental concern: to what extent should one trust an IDS?~\cite{elhouda2022idstrustworthiness}. This issue has led to the development of eXplainable Artificial Intelligence (XAI) approaches to improve the trustworthiness and effectiveness of IDSs~\cite{nascita2024ntaxaisurvey}. Most existing XAI methods for IDSs rely on post-hoc, feature-based explanations, employing frameworks such as SHAP and LIME~\cite{hariharan2023xai}. However, in the absence of carefully designed feature engineering, these explanations are inherently limited to quantifying feature contributions to a given prediction and remain unable to convey richer information, such as attack dynamics and packet-level sequencing. Consequently, such explanations lack a true process-based view of the underlying malicious behavior.

In this paper, we investigate the use of process mining to \textit{enhance} the diagnoses provided by anomaly-based IDSs in terms of alarm severity rating and explanation. Process mining is a research area aimed at bridging data-driven insights and process science~\cite{aalst2016pm}, and has been employed to analyze several network protocols in different application scenarios~\cite{ahmadon2020cpsmqttpbad, hornsteiner2024pmopcua, vitale2025networktrafficanalysisprocess}. \revision{Its utility mainly lies in extracting process models that can capture control-flow relations between the packets exchanged in networked communications. Our method leverages the capabilities of process mining for process-based explanation and rating of anomaly-based IDS alarms. This way, we can prioritize critical alerts and maintain visibility into network behavior, while minimizing disruption by allowing misclassified benign traffic to pass.}

The proposed method involves two phases: 1) the training phase, in which a reference anomaly-based IDS is trained with normal network flows and validated against anomalous network flows, and a packet-level process-based characterization of false-positive network flows is built through process mining; and 2) the inference phase, in which alarms raised by the IDS are examined against the process model. Here, the method provides a severity rating inversely related to the similarity with the false-positive process-based explanation (i.e., high similarity indicates a likely false positive and thus low severity). Thus, the proposal operates as an example-based XAI approach~\cite{nascita2024ntaxaisurvey}, in which false-positive flows are recognized and discriminated from high-severity alarms. Since our approach can be applied to any black-box model and does not rely on model-specific characteristics, it can be configured as a model-agnostic XAI technique.

In summary, the proposed method brings the following novelties:
\begin{itemize}
    \item A severity rating mechanism of anomaly-based IDS alarms that utilizes the packet-level process-based explanations to discriminate high-severity true positives from false positive traffic.
    \item Novel explanation mechanism for anomaly-based IDS alarms based on process models mined from network flows.
\end{itemize}

We experiment with our approach using the publicly available USB-IDS-TC dataset~\cite{icissp25}. The dataset includes instances of normal traffic flows and traffic flows affected by different variants of the Slowloris DoS attacks. The results show that our method is able to discriminate between low- to very-high-severity alarms while achieving 99.94\% recall and 99.99\% precision, effectively discarding false positives while providing different degrees of severity for the true positives.

%The rest of the paper is organized as follows. Section \ref{sec:sota} reviews existing work on XAI, IDSs, and process mining for IDSs; Section \ref{sec:method} provides a detailed description of our approach; Section \ref{sec:evaluation} describes the datasets used and the results obtained; and Section \ref{sec:conclusion} concludes and presents future work.
\section{Related Work}
\label{sec:sota}
\subsection{Explainable Artificial Intelligence for Intrusion Detection Systems}
%XAI is defined as “AI systems that can explain their rationale to human users, characterize their strengths and weaknesses, and convey an understanding of their future behavior”~\cite{gunning2019darpa}. 
The utility of XAI, as well as the challenges associated with its adoption, has been widely reported across both scientific and industrial contexts. Nascita et al.~\cite{nascita2024ntaxaisurvey} surveyed the use of XAI in network traffic classification and intrusion detection, categorizing existing approaches according to their scope, explanation stage, model dependency, and explanation type. In particular, they highlight the role of XAI in enhancing confidence in IDSs by providing actionable insights through alarm rating.

While this categorization opens up several opportunities for XAI adoption, the existing literature has largely focused on SHAP and LIME. These techniques have been successfully applied to a wide range of attack types, including DNS-over-HTTPS attacks~\cite{zebin2022idsxai} and DoS attacks~\cite{hariharan2023xai}. However, these approaches treat features statically and often struggle to capture the temporal dynamics and sequential dependencies inherent in network traffic. Consequently, they can produce inconsistent or partially conflicting explanations, prompting recent studies to emphasize the need for combining multiple XAI approaches to improve explanation robustness and user confidence~\cite{dai2025explainableidsiot}.

\subsection{Process Mining for \revision{Explaining} Intrusion Detection Systems}
%Process mining is a hybrid discipline whose objective is “to discover, monitor, and improve real processes by extracting knowledge from event logs”~\cite{aalst2016pm}. The main types of process mining involve building process models (process discovery) and checking whether new behavior aligns with process-based characterizations (conformance checking). 
Process mining is a hybrid discipline whose main tasks involve building process models (process discovery) and checking whether new behavior aligns with process-based characterizations (conformance checking). This is very useful for \revision{enhancing} intrusion detection, as it offers means to capture network patterns and verify deviations in an explainable manner. Specifically, process mining has been applied to inspect normal DNS traffic~\cite{saintpierre2014dnspmad}, analyzing MQTT message patterns~\cite{ahmadon2020cpsmqttpbad}, and studying TCP traffic~\cite{vitale2025networktrafficanalysisprocess}.

\revision{Process mining has been used for the explanation of control-flow anomalies through fine-tuning of sequence-to-sequence language models~\cite{busch2024xsemad} and the extraction of explainable process models of industrial control systems~\cite{vitale2025pmdt}.} Notably, process mining has been used as an additional layer of explainability in combination with intrusion detectors. De Alvarenga et al.~\cite{dealvarenga2018pmhc} proposed process discovery as a means to model attackers' behavior and provide further insight and confidence in alerts raised by IDSs. Wang et al.~\cite{wang2022lstmaeaddiagnosis} enhanced intrusion detectors with explainable post-processing of IDS alerts through process mining. Their method involved capturing normal traffic behavior and checking the traffic flagged as anomalous against such characterization to outline the process-based deviations. \revision{However, neither approach integrated a quantitative rating mechanism to systematically filter or rank alarms based on severity.}

\subsection{Anomaly Rating}

\revision{Alarm rating mechanisms are particularly beneficial for security analysts in mitigating alert fatigue. An approach is discussed in ~\cite{malach2025cybershapley}, where sequences of log events are represented as graphs and prioritized according to their associated Shapley values. Himmelhuber et al. ~\cite{himmelhuber2022detection} proposed a methodology that combines GNNExplainer with DL-Learner to filter and prioritize alerts leveraging fidelity scores within Industrial Control Systems. In an IIoT environment, Khalaf et al. ~\cite{khalaf2025real} prioritized alerts based on an adaptive threat correlation engine leveraging attention-weighted graph structures and sequence modeling.} 

\revision{While the methods above provide insights into alerts and their rating, they did not employ process-based explanations to rate alarms to uncover anomalous network patterns}. Instead, we integrate process mining to enhance intrusion detection through process-based explanation and ranking of anomaly-based IDS alerts. Not only does our method provide insight into network processes, but it also offers a tunable alert rating system that enables selective filtering of anomalous traffic.
\section{Method}
\label{sec:method}
%\note{tagiliare cappello e titolo prima sezione}
%In this section, we provide an introduction to the problem statement addressed by our method and a detailed description of the training and inference phases.

%\subsection{Problem Statement}
\begin{figure*}[!t]
    \centering
    \includegraphics[width=0.8\textwidth]{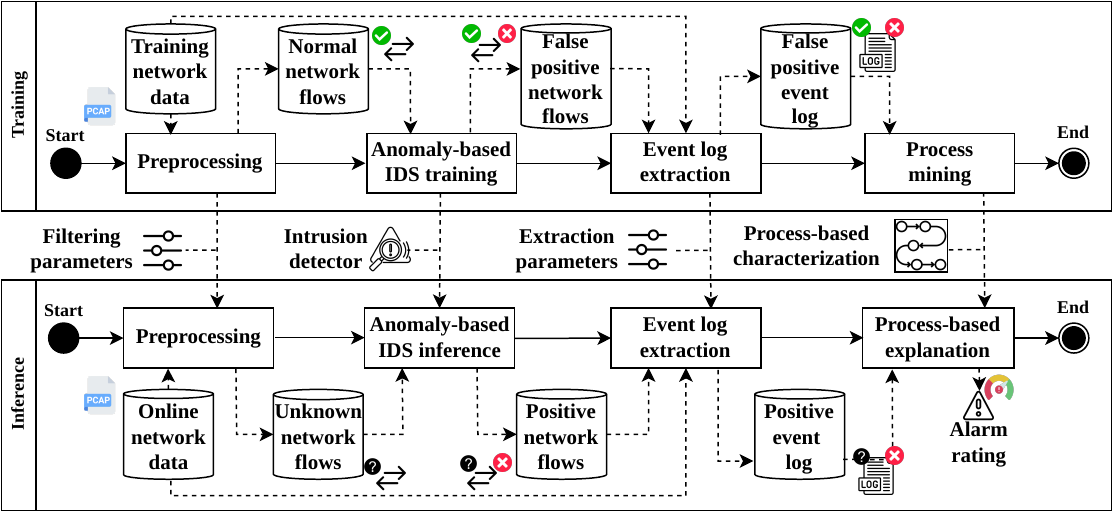}
    \caption{The proposed method for process-based explanation and ranking of anomaly-based IDS alerts. The first phase involves training an anomaly-based IDS and mining a process-based characterization of false positive network flows. The second inference phase classifies incoming network flows through the IDS and rates the positive network flows.}
    \label{fig:framework}
\end{figure*}

%\begin{figure*}[!t]
%\centering
%\begin{subfigure}{0.9\columnwidth}
%\includegraphics[width=\columnwidth]{figures/training.png}%
%\caption{Training process.}%
%\label{fig:train}%
%\end{subfigure}\hfill%
%\begin{subfigure}{0.9\columnwidth}
%\includegraphics[width=\columnwidth]{figures/inference.png}%
%\caption{Inference process.}%
%\label{fig:inference}%
%\end{subfigure}\hfill%
%\caption{Enhanced intrusion detection process.}
%\label{fig:process}
%\end{figure*}

%Anomaly-based IDSs employ approaches to learn normal network traffic and identify any deviating traffic that diverges from it, provided a specified upper limit of similarity is met. This allows relaxing the constraint of having labeled data and enables the detection of zero-day attacks \cite{segurola2024unsupervised}. On the other hand, several false positives may occur if the training data are contaminated with noisy samples \cite{hozouri2025comprehensive}, and the black-box nature of deep learning-based classifiers makes it difficult to understand alarm severities~\cite{nascita2024ntaxaisurvey}.  
%Our proposal, depicted in Figure \ref{fig:framework}, aims to mitigate the aforementioned anomaly-based IDS drawbacks through process-based explanations and ratings of alarms. 
Anomaly-based IDSs employ approaches to learn normal network traffic and identify any deviating traffic that diverges from it, provided a specified upper limit of similarity is met. However, several false positives may occur if the training data are contaminated with noisy samples \cite{hozouri2025comprehensive}, and the black-box nature of deep learning-based classifiers makes it difficult to understand alarm severities~\cite{nascita2024ntaxaisurvey}.  

Our proposal, depicted in Figure~\ref{fig:framework}, builds on standard anomaly-based IDSs and introduces a process mining layer to provide process-based explanations of anomalous network traffic and rate alarms. \revision{While our approach can be applied to various protocols, we focus on a network dataset based on TCP traffic, described later in detail.} In the following, we detail the training and inference phases with reference to the \textit{anomaly-based IDS pipeline} and \textit{process mining for alarm rating and explanation}. 
%The former focuses on the data processing and learning stages that precede the integration of the process mining layer, whereas the latter details how process mining is used to enhance anomaly-based IDSs with packet-level process-based analyses.

\subsection{Anomaly-Based IDS Pipeline}

\subsubsection{Training}
Firstly, the method involves \textbf{preprocessing} training network data to extract network flows corresponding to specific protocols prior to the training and inference phases. A network flow is defined by the tuple described by the source and destination IP addresses and port numbers, and it is characterized by the network data exchanged in a bidirectional connection between two hosts. 
During preprocessing, the method extracts \textit{normal network flows} from raw PCAPs. For each PCAP, the network flows are extracted using well-known tools, such as CICFlowMeter \cite{Lashkari2017253} and NTLFlowLyzer \cite{shafi_ntlflowlyzer_2025}. It is possible to group the network packets based on the computed flows to which they belong; in this way, both the flows and the related PCAPs are obtained.

Normal network flows can be used for \textbf{anomaly-based IDS training}. This step allows building an anomaly-based IDS that produces binary classifications, in which negative outcomes correspond to network flows considered normal, while positive outcomes indicate anomalous flows and thus represent alerts. The training process of anomaly-based IDSs involves learning normal behavior from benign network flows using, for example, one-class classification or reconstruction-based techniques. This process requires the selection of a threshold on a decision metric used to classify previously unseen network flows. A strict threshold increases the number of false positives, whereas a relaxed threshold increases the number of false negatives. The threshold can be determined empirically using a validation set, which enables the selection of a conservative value that accounts for variability in normal behavior and improves robustness at inference time. However, the validation process almost always leads to a certain number of \textit{false positive network flows}, which represent benign traffic that is mistakenly classified as anomalous. 
\revision{Since the objective of the approach is to evaluate alarms based on false positives, a more restrictive threshold can be applied to increase the number of false-positive network flows, thereby making the subsequent process mining-based modeling more robust.}

\subsubsection{Inference}
Once the anomaly detection model and its associated thresholds have been defined, the IDS can be deployed for continuous networking monitoring. \textit{Online network data} is captured, organized in PCAPs, and subject to \textbf{preprocessing} according to the \textit{filtering parameters} obtained at training time to construct \textit{unknown network flows}. These network flows are run through \textbf{anomaly-based IDS inference} with the \textit{intrusion detector} built at training time. The inference process leads to both negative and positive network flows. Since the goal of the approach is to rate alarms based on their severity, only the \textit{positive network flows} are used for further processing.

\subsection{Process Mining for Alarm Rating and Explanation}
\subsubsection{Training}We adopt the process mining-based network traffic analysis framework proposed in~\cite{vitale2025networktrafficanalysisprocess}, which first involves \textbf{event log extraction} from packet-level PCAP data corresponding to the false positive network flows. The framework aims to identify distinct TCP network event patterns, referred to as states, by applying an unsupervised clustering-based approach. Formally, given the set of TCP flags $\Sigma$,  $\Sigma^*$ the universe of sequences that can be built with the TCP flags, $\mathcal{B}(\Sigma^*)$ the universe of bags that can be built over $\Sigma^*$, $\Xi$ the set of states obtained from the false-positive PCAP data, and $\xi\in\Xi$ a state obtained through the clustering-based approach, an event log $L_{\xi}\in\mathcal{B}(\Sigma^*)$ is a bag of TCP traces belonging to state $\xi$. Each trace $\sigma_{\xi}\in L_{\xi}$ is associated with the part of the network flow that belongs to that state. This implies that a network flow can have multiple states, captured within different event logs.

\begin{figure}[!t]
    \centering
    \includegraphics[width=\columnwidth]{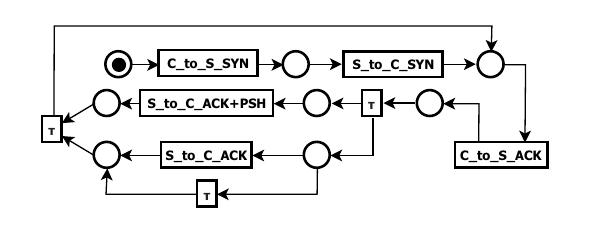}
    \caption{Example of a Petri net capturing a TCP event flow pattern.}
    \label{fig:example_pn}
\end{figure}

Let $\mathcal{L}=\{L_{\xi}\in\mathcal{B}(\Sigma^*):\xi\in\Xi\}$ be the set of \textit{false-positive event logs} obtained from the false-positive PCAP data at training time. The \textbf{process mining} step builds $|\Xi|$ process models through a process-discovery algorithm. Specifically, given algorithm $\gamma$ and $L_{\xi}\in\mathcal{L}$, $N_{\xi}=\gamma(L_{\xi})$ is the process model that captures the sequences of TCP events in state $\xi$. Usually, $\gamma$ extracts a Petri net~\cite{aalst2016pm}. Fig. \ref{fig:example_pn} shows an example Petri net capturing a TCP event flow pattern. The squared nodes indicate transitions, which can either be TCP events (e.g., C\_to\_S\_SYN) or silent ($\tau$) events. The rounded places constrain the allowed sequencing of the transitions. The place with a full circle in it indicates the ``initial marking'', i.e., the presence of a single token in a ``start'' place allowing the execution of C\_to\_S\_SYN. The semantics of the Petri net involve moving tokens across the places, starting from the initial marking. For example, the execution of C\_to\_S\_SYN moves the token to the next place, allowing the execution of S\_to\_C\_SYN.
\revision{The quality of the $|\Xi|$ Petri nets obtained from false-positive event logs depends on the \textit{completeness} of the logs, meaning the availability of enough examples to identify control-flow relations among TCP events. In the presence of incomplete logs, one can estimate the probability of modeling a control-flow relation based on its observed frequency~\cite{leemans2014discovering}.}

In the following, we denote $\mathcal{N}$ the set of Petri nets obtained from event logs of $\mathcal{L}$, i.e., the \textbf{process-based characterization}.
$N_{\xi}\in\mathcal{N}$ can be used to check whether an event log $L_{\xi}\in\mathcal{B}(\Sigma^*)$ conforms to the expected TCP event patterns. In particular, we consider the class of alignment-based conformance checking algorithms, which allow extracting the so-called alignments. Let us consider an illustrative alignment example with the Petri net in Fig. \ref{fig:example_pn} and a trace $\sigma=\langle$C\_to\_S\_SYN, S\_to\_C\_SYN, S\_to\_C\_ACK+PSH, C\_to\_S\_ACK$\rangle$. An aligned trace $\sigma_A$ attempts to find a path through the model that optimally matches $\sigma$. In this case, the first two TCP events are perfectly matched to the Petri net, but the subsequent S\_to\_C\_ACK+PSH cannot be executed without the prior execution of C\_to\_S\_ACK. In this case, $\sigma_A$ should account for this mismatch. For example, $\sigma_A=\langle$C\_to\_S\_SYN, S\_to\_C\_SYN, C\_to\_S\_ACK, S\_to\_C\_ACK+PSH, C\_to\_S\_ACK$\rangle$ matches all the TCP events except for C\_to\_S\_ACK after S\_to\_C\_SYN. In this paper, we aim to count the number of alignments associated with each TCP event of a network traffic flow, similar to the approach described in~\cite{vitale2025cfad}. In particular, we build the set of alignments associated with each TCP event $\mathcal{A}=\{A_{t}\in\mathbb{N}:t\in\Sigma\}$ by performing alignment-based conformance checking between the Petri nets of $\mathcal{N}$ and the traces of the event logs of $\mathcal{L}$, where $A_{t}$ is averaged across all the traces. $\mathcal{A}$ will be subsequently used in the inference phase to rate the alerts.

%\note{daMax: vedo qualche criticità, la descrizione va subito nel dettaglio, ma non si capisce l'idea generale. Inoltre per chi non conosce l'idea non si coglie il fatto che da una prte c'è lapproccio ML classico, e poi in parallelo aggiungiamo il process mining }

%\note{Grims: le parti di preprocessing e IDS inference potrebbero necessitare di modifiche.}

\subsubsection{Inference}Following the same process as in the training phase, the positive network flows are run through \textbf{event log extraction} with the same \textit{extraction parameters} of the mentioned clustering-based approach. This way, the set of positive event logs $\mathcal{L}_{p}$ are used for \textbf{process-based explanation} of the alarms. Let us denote $\sigma_{\xi}\subseteq\sigma$ the subtrace of trace $\sigma$ that belongs to state $\xi$. An individual network flow $f$ to which a trace $\sigma_f$ is associated is decomposed into multiple subtraces belonging to the different states of $\Xi$: $f=\{\sigma_{f,\xi}\subseteq\sigma_f:\xi\in\Xi\}$. The process-based explanation of the network flow $f$ is the set of alignments $\mathcal{A}_{f}=\{A_{t,f}\in\mathbb{N}:t\in\Sigma\}$ obtained by performing alignment-based conformance checking between each subtrace of $f$ and the corresponding Petri net. Finally, we compute the \textit{alarm rating} of $f$ by evaluating the similarity between $\mathcal{A}_{f}$ and $\mathcal{A}$. In particular, we use the cosine similarity:
\begin{equation*}
    \textrm{CosSim}(\mathcal{A},\mathcal{A}_f)=\frac{\langle\mathcal{A},\mathcal{A}_f\rangle}{||\mathcal{A}||\,||\mathcal{A}_f||}.
\end{equation*}
The lower the similarity between $\mathcal{A}$ and $\mathcal{A}_f$, the more severe the alarm is; in contrast, the higher the similarity, the closer the alarm is to a false positive.

\section{Evaluation}
\label{sec:evaluation}
%In this section, we aim to demonstrate the effectiveness of our approach at discriminating different alert severities while maintaining high classification performance. We first introduce the dataset properties and the attacks, review the techniques and metrics to evaluate the two goals, and report and discuss the results obtained.

\subsection{Dataset}
In our research, we utilized the publicly available USB-IDS-TC dataset, which was designed to address the issue of the dependence of machine learning- and deep learning-based network IDS on the network scenario used to collect training traffic data \cite{icissp25}. The testbed, described in detail in the referenced work, is built on Docker containers, and includes a web server, an attacker and a tester used both to issue randomized web requests and collect availability statistics. 
%The dataset was initially organized in four separate PCAPs, one related to normal traffic and the remaining ones to three different types of DoS attacks: slow post and two kinds of slow loris. More accurate information about PCAPs is summarized in Table \ref{tab:pcaps}. The normal traffic (NOR) present an higher number of packets and, as expected, a lower packets-per-second rate compared to slow loris (GSL, via \textit{slowloris\footnote{https://github.com/gkbrk/slowloris}}, high-intensity slow loris (HSL, via \textit{slowhttptest} in slow loris mode), and high-intensity slow post (HSP, via \textit{slowhttptest} in slow post mode) DoS attacks. 
The dataset was initially organized in four separate PCAPs, one related to normal traffic (NOR) and the remaining ones to three different types of DoS attacks: slow loris (GSL) via \textit{slowloris\footnote{\url{https://github.com/gkbrk/slowloris}}}, high-intensity slow loris (HSL), and slow post (HSP) via \textit{slowhttptest\footnote{\url{https://www.kali.org/tools/slowhttptest/}}}. Classified as \textit{low and slow}, these DoS attacks saturate TCP connections of the system under attack through traffic that appears legitimate. \revision{Network traffic generated by such attacks is characterized by an increased volume of TCP packets featuring [PSH, ACK] flags, which are distributed over time by the malicious host. This leads to an alteration in the network traffic control flow, which consequently causes misalignments in the TCP traces modeled in the process-based characterization.}

%Their distinctive effect lies in altering the casual structure of TCP events observed in normal traffic, which out process mining-based approach effectively identifies.
%More accurate information about PCAPs is summarized in Table \ref{tab:pcaps}. The normal traffic presents a higher number of packets and, as expected, a lower packets-per-second rate compared to DoS attacks. 
% NOR: normal
% GSL: slowloris 
% HSL: slowhttptest in slow loris mode
% HSP: slowhttptest in slow post mode
%\begin{table}
%\input{tables/pcap_info}
%\end{table}

As described in Section \ref{sec:method}, the detection model is trained and evaluated on network flows. For this purpose, bidirectional network flows were extracted and organized in CSV format from PCAPs data using NTLFlowLyzer. The tool allows for the retrieval of 348 features, enhancing and expanding the capabilities of CICFlowMeter, and a summary of the statistics is shown in Table \ref{tab:flows}, with counts reported as absolute values and all other metrics expressed as means with standard deviations. It is worth noting that, although the total number of normal traffic packets is significantly higher, the number of flows generated from them is substantially lower. In addition, an interesting metric is the \textit{flows-per-minute}, as there is a difference of two orders of magnitude between the values observed for normal traffic and those for anomalous traffic. This is due to the nature of the traffic itself and to the specific characteristics of the executed attacks. 

Following an initial preprocessing step to remove incomplete network flows, the dataset was partitioned to enable the training and testing of the three: One-Class SVM (OCSVM), Autoencoder, and Variational Autoencoder. In particular, 969 normal network flows were used to train the OCSVM, while 833 were employed for the autoencoders. In both scenarios, normal network flows were randomly partitioned into training and validation sets five times. This procedure ensures five independent runs of the proposed method, effectively accounting for the stochastic nature of IDS training. \revision{At each run, the misclassified validation normal flows are used to extract the process-based characterization. The different process-based characterizations across the five runs account for the possibility of overfitting on restricted sets of normal network flows.}
The remaining 136 flows were reserved for the validation phase to compute the decision threshold based on the reconstruction error. Finally, the remaining 647 normal network flows were combined with anomalous flows and used in the testing phase, for a total of 14869 flows.

\begin{table}[!t]
    \begin{center}
    \caption{Mean and standard deviation of different network traffic flow metrics.}
    \label{tab:flows}
    \resizebox{\columnwidth}{!}{%
    \begin{tabular}{lcccc}
        \toprule
        \textbf{Metric} & \textbf{NOR} & \textbf{GSL} & \textbf{HSL} & \textbf{HSP} \\
        \midrule
        N. flows & 1616 & 4052 & 5088 & 5082 \\
        Flows-per-minute & 76.1 & 1403.3 & 1774.9 & 1732.5 \\
        Flow duration (s) & 57$\pm$49 & 37$\pm$17 & 60$\pm$27 & 61$\pm$30 \\
        Bytes-per-flow & 322388$\pm$568000 & 956$\pm$261 & 1589$\pm$696 & 2204$\pm$1290 \\
        Packets-per-flow & 182$\pm$292 & 14$\pm$3 & 19$\pm$6 & 20$\pm$6 \\
        Packet length (byte) & 1389$\pm$427 & 69$\pm$8 & 80$\pm$25 & 108$\pm$53 \\
        \midrule
        ACK flags & 181$\pm$292 & 12$\pm$3 & 14$\pm$6 & 15$\pm$6 \\
        SYN flags & 2$\pm$0.2 & 2$\pm$0.3 & 5.8$\pm$3.7 & 5.8$\pm$3.7 \\
        FIN flags & 2$\pm$0.2 & 1$\pm$0.4 & 1.3$\pm$1.0 & 1.3$\pm$1.0 \\
        RST flags & 0.02$\pm$0.2 & 0.8$\pm$0.4 & 0.1$\pm$0.3 & 0.1$\pm$0.3 \\
        PSH flags & 101$\pm$168 & 4$\pm$1 & 6.9$\pm$3.7 & 7.1$\pm$3.7 \\
        \midrule
        Forward packets & 61$\pm$89 & 7$\pm$2 & 13$\pm$5 & 13$\pm$5 \\
        Backward packets & 121$\pm$204 & 6$\pm$2 & 6$\pm$4 & 7$\pm$4 \\
        Forward bytes & 307$\pm$226 & 195$\pm$36 & 652$\pm$626 & 1187$\pm$1240 \\
        Backward bytes & 316221$\pm$558074 & 320$\pm$165 & 274$\pm$242 & 347$\pm$518 \\
        \bottomrule
    \end{tabular}

}
    \end{center}
\end{table}

\subsection{Techniques, Metrics, and Software} 
The three techniques used for anomaly-based IDS training are three popular one-class classification and reconstruction-based approaches: One-Class Support Vector Machine (OCSVM), the Autoencoder (AE), and the Variational Autoencoder (VAE). The AE and VAE adopt a multilayer architecture with the hidden layers with 400 and 200 units, ReLU as activation function, and are trained minimizing the mean squared error loss with a learning rate of 0.001, batch size 64, and 150 epochs. In addition, the VAE includes a latent space of dimension 6. The OCSVM adopt RBF kernel to handle non-linear relationships, with the parameter $\nu$ set to 0.15. All configurations are chosen according to~\cite{paolini2025one}, except for the OCSVM $\nu$ parameter, which was empirically tuned.

Regarding event log extraction and process mining. We configured event log extraction with k-means clustering by setting k equal to 2 and a sliding window length equal to 3. The configuration of two network states allows the discrimination of different TCP patterns without isolating those that appear in both normal and anomalous traffic, which would make their separation more challenging. On the other hand, a window length equal to 3 allows reducing the underfitting effect documented in~\cite{vitale2025networktrafficanalysisprocess}. As for process discovery, we apply the vanilla inductive miner to each state-wise event log. 

To define the severity bands, we use the CosSim metric, which takes values in the range $[0, 1]$. The severity bands are defined by partitioning this range into five intervals based on predefined thresholds: $[0, 0.01)$ (Very High), $[0.01, 0.25)$ (High), $[0.25, 0.75)$ (Medium), $[0.75, 0.99)$ (Low), and $[0.99, 1.00]$ (Very Low). This results in the set of severity bands $\mathcal{S}=\{\textrm{Very High}, \textrm{High}, \textrm{Medium}, \textrm{Low}, \textrm{Very Low}\}$. \revision{These bands are defined so that obvious alarms (Very High and High) can be filtered without impacting normal traffic, while uncertain alarms (Low and Very Low) are neglected to retain potentially misclassified normal traffic. The Medium band, which is the largest, captures uncertainty and may be further subdivided when distinguishing malicious from benign traffic is challenging.}

To evaluate the effect of splitting alerts into different severity bands, we introduce modified versions of the recall and precision metrics. Specifically, for a given severity threshold indexed by $k \in \{1, \dots, N\}$, where $N = |\mathcal{S}|$, we evaluate the impact of filtering traffic by discarding alerts belonging to severity bands higher than $k$ (i.e., treating them as normal traffic):
\begin{flalign*}
    \textrm{Recall}_{k}&=\frac{\sum_{i=1}^{k} \textrm{TP}_i}{\sum_{i=1}^{k}\textrm{TP}_{i}+\sum_{i=k+1}^{N}\textrm{TP}_i+\textrm{FN}},&\\
    \textrm{Precision}_{k}&=\frac{\sum_{i=1}^{k} \textrm{TP}_i}{\sum_{i=1}^{k} \textrm{TP}_i + \sum_{i=1}^{k}\textrm{FP}_i}.
\end{flalign*}
TP$_i$ and FP$_{i}$ indicate the true and false positives in the $i$-th severity band, whereas FN indicates the global false negatives.

The implementation of the proposed method is publicly available on GitHub\footnote{\url{https://github.com/francescovitale/pm_based_ids_rating}}. The software is developed in Python and was executed on a Windows 11 workstation equipped with an Intel® Core™ i9-11900K CPU @ 3.50GHz and 32GB of RAM. The framework leverages standard machine learning and process mining libraries, specifically \texttt{scikit-learn} and \texttt{pm4py}.

\subsection{Results}
\begin{figure}[!t]
    \centering
    \includegraphics[width=0.9\columnwidth]{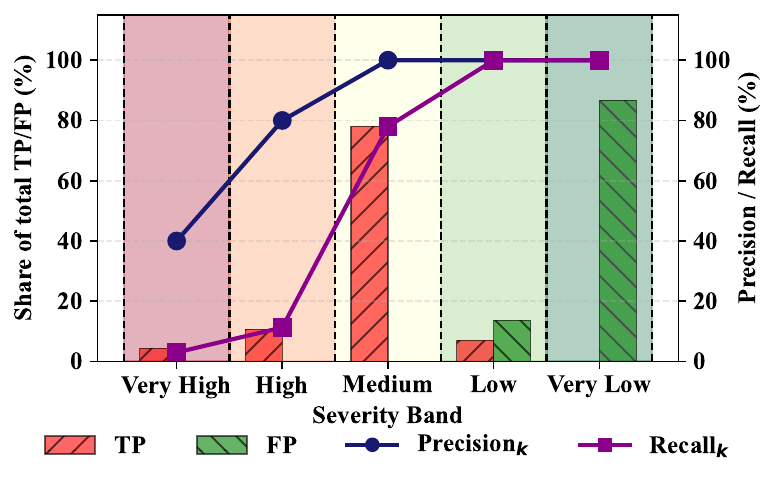}
    \caption{The recall and precision metrics, together with the share of total TP and FP percentages, across the different severity bands, related to the AE classifier.}
    \label{fig:performance}
\end{figure}

Table \ref{tab:rating_results} shows the performance in terms of TP, FP, recall and precision of the anomaly-based models for each severity band. As expected, discarding alarms from the high, medium, low and very-low severity bands leads to very low recall, achieving only up to 4.44\% with the AE. This is because only a small number of alarms fall into the very-high severity band. To improve the overall performance, lower severity bands should be included to increase the number of alarms being analyzed. For all the models, placing in the low severity band allows including the majority of true positives while discarding many false positives. For example, by including all alarms down to the low severity band and discarding the alarms in the very-low severity band, the AE peaks at 99.94\% recall and 99.99\% precision by covering on average 14141 TPs and 2 FPs as alarms, and considering 8 TPs and 8 FPs as benign traffic. The recall and precision metrics, together with the total share of TP and FP percentages, across the different severity bands, related to the AE classifier, are shown in Fig. \ref{fig:performance}. This visualization highlights that most TPs fall within the very high, high, and medium severity bands, whereas FPs are primarily found in the low and very low severity bands.

\begin{table}[!t]
\centering
\caption{The performance of the anomaly-based models for each severity band. \revision{TP$_{k}$ and FP$_{k}$ represent the isolated alerts within that specific band, whereas Recall$_{k}$ and Precision$_{k}$ are cumulative metrics evaluated by passing all traffic from the most severe band ($k=1$) up to the current threshold $k$.}}
\label{tab:rating_results}
\resizebox{\columnwidth}{!}{%
\begin{tabular}{llllll}\toprule
\textbf{Severity} & \textbf{Model} & \textbf{TP$_{k}$} & \textbf{FP$_{k}$} & \textbf{Recall$_{k}$ (\%)} & \textbf{Precision$_{k}$ (\%)} \\ \midrule
\multirow{3}{*}{\shortstack[l]{Very low \\ ($k=5$)}} & OCSVM & 4$\pm$9 & 25$\pm$3 & 100.00$\pm$0.00 & 99.67$\pm$0.01 \\
 & AE & 8$\pm$11 & 8$\pm$1 & 100.00$\pm$0.00 & 99.94$\pm$0.01 \\
 & VAE & 8$\pm$11 & 11$\pm$3 & 100.00$\pm$0.00 & 99.91$\pm$0.00 \\
\midrule
\multirow{3}{*}{\shortstack[l]{Low \\ ($k=4$)}} & OCSVM & 81$\pm$158 & 22$\pm$4 & 99.97$\pm$0.06 & 99.85$\pm$0.03 \\
 & AE & 993$\pm$1209 & 2$\pm$2 & 99.94$\pm$0.08 & 99.99$\pm$0.01 \\
 & VAE & 2460$\pm$403 & 3$\pm$3 & 99.94$\pm$0.08 & 99.98$\pm$0.01 \\
\midrule
\multirow{3}{*}{\shortstack[l]{Medium \\ ($k=3$)}} & OCSVM & 11184$\pm$1387 & 1$\pm$1 & 99.40$\pm$1.18 & 100.00$\pm$0.00 \\
 & AE & 11036$\pm$1981 & 0$\pm$0 & 92.93$\pm$8.56 & 100.00$\pm$0.00 \\
 & VAE & 10882$\pm$946 & 0$\pm$0 & 82.56$\pm$2.77 & 100.00$\pm$0.00 \\
\midrule
\multirow{3}{*}{\shortstack[l]{High \\ ($k=2$)}} & OCSVM & 2263$\pm$1274 & 0$\pm$0 & 20.35$\pm$10.97 & 100.00$\pm$0.00 \\
 & AE & 1484$\pm$1308 & 0$\pm$0 & 14.92$\pm$13.26 & 80.00$\pm$44.72 \\
 & VAE & 586$\pm$612 & 0$\pm$0 & 5.64$\pm$4.02 & 80.00$\pm$44.72 \\
\midrule
\multirow{3}{*}{\shortstack[l]{Very high \\ ($k=1$)}} & OCSVM & 616$\pm$563 & 0$\pm$0 & 4.35$\pm$3.97 & 60.00$\pm$54.77 \\
 & AE & 628$\pm$574 & 0$\pm$0 & 4.44$\pm$4.05 & 60.00$\pm$54.77 \\
 & VAE & 212$\pm$474 & 0$\pm$0 & 1.50$\pm$3.35 & 20.00$\pm$44.72 \\
\bottomrule
\end{tabular}
}
\end{table}

\begin{figure}[!t]
    \centering
    \includegraphics[width=0.9\columnwidth]{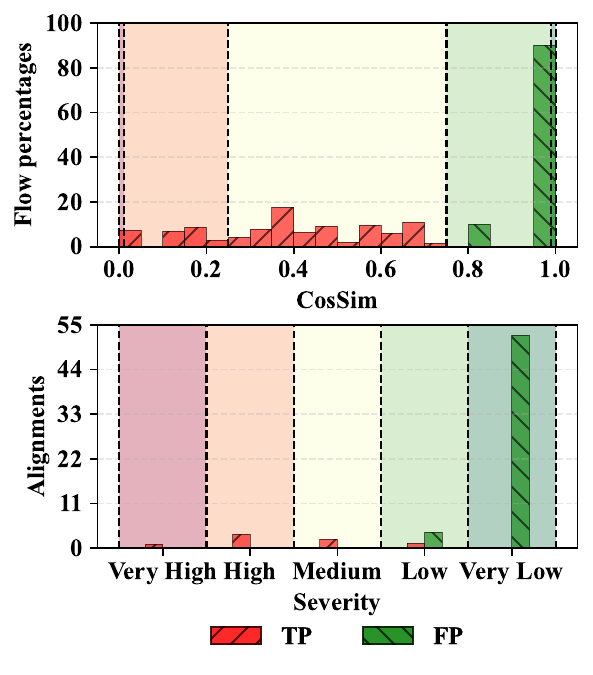}
    \caption{The percentages of TP and FP network flows falling in each severity band (top) and the mean number of alignments for each band associated with the TPs and FPs falling into the severity bands (bottom), related to the AE classifier.}
    \label{fig:explanation}
\end{figure}

The interpretation of the results is backed by the bar plots shown in Fig. \ref{fig:explanation}. At the bottom is the process-based explanation in terms of the mean number of alignments for each band associated with the TPs and FPs falling into the severity bands. This visualization outlines that TP network flows are much less similar to the process-based characterization obtained in the training phase, whereas FP network flows, especially those falling in the very low severity band, are much more aligned, hence more likely to be misclassified. The bar plots at the top show the percentages of network flows falling in each severity band according to their CosSim. As expected, TP network flows have mostly CosSim figures from 0.0 to 0.75 (very-high to medium severity), whereas FP network flows have mostly CosSim figures from 0.75 to 1.00 (low to very-low severity).

In conclusion, the results highlight that the proposed method for process-based rating and explanation of anomaly-based IDS alerts shows consistent and coherent performance across the classifiers, allowing the severity labeling of each alarm into discriminative bands.

\section{Conclusion}
\label{sec:conclusion}

While anomaly-based IDSs employing machine and deep learning techniques yield outstanding results, their trustworthiness remains a critical challenge. Despite numerous XAI proposals, the literature lacks process-based explanations that connect IDS decisions to the actual sequencing of network packets. To address this gap, we proposed a method for process-based explanation and rating of IDS alerts.
%, which enables rating anomalous network flows across different severity bands. The method 
%leverages residual false-positive network flows identified during anomaly-based IDS training to 
The method builds a process-based characterization of false-positive network flows using process mining techniques. During inference, IDS alerts are compared against this characterization to evaluate their similarity to false positives, effectively distinguishing high-severity traffic from likely false alarms. We validated our approach using the USB-IDS-TC dataset, focusing on various Slowloris DoS attack variants. Results demonstrated that our method successfully discriminates between low- and very-high-severity alarms while maintaining high detection performance.

In future works, we plan to: 1) expand our method by integrating automatic tuning of state-space characterization to improve the distinction between false positives and actual malicious behavior; \revision{2) extend the sensitivity analysis to a broader set of process discovery algorithms and severity band thresholds; 3) sharpen the process-based explanations with personalized costs for specific misaligned TCP events; 4) comparison of our results with other alarm ranking methods;} 5) extend experimentation to other relevant attack datasets to strengthen the generalizability of our process-based explanation; and 6) investigate other network protocols across the ISO/OSI stack.

%\section*{Acknowledgment}

\bibliographystyle{IEEEtran}
\bibliography{bibliography}

@article{khraisat2019survey,
  title={{Survey of intrusion detection systems: techniques, datasets and challenges}},
  author={Khraisat, Ansam and Gondal, Iqbal and Vamplew, Peter and Kamruzzaman, Joarder},
  journal={Cybersecurity},
  volume={2},
  number={1},
  pages={1--22},
  year={2019},
  publisher={Springer}
}

@article{dealvarenga2018pmhc,
  title={{Process mining and hierarchical clustering to help intrusion alert visualization}},
  author={De Alvarenga, Sean Carlisto and Barbon Jr, Sylvio and Miani, Rodrigo Sanches and Cukier, Michel and Zarpel{\~a}o, Bruno Bogaz},
  journal={Computers \& Security},
  volume={73},
  pages={474--491},
  year={2018},
  publisher={Elsevier}
}

@inproceedings{wang2022lstmaeaddiagnosis,
author = {Wang, Xiaolei and Yang, Lin and Li, Dongyang and Ma, Linru and He, Yongzhong and Xiao, Junchao and Liu, Jiyuan and Yang, Yuexiang},
title = {{MADDC: Multi-Scale Anomaly Detection, Diagnosis and Correction for Discrete Event Logs}},
year = {2022},
doi = {10.1145/3564625.3567972},
booktitle = {Proceedings of the 38th Annual Computer Security Applications Conference},
pages = {769–784},
numpages = {16}
}

@inproceedings{saintpierre2014dnspmad,
  title={{Detecting anomalies in DNS protocol traces via passive testing and process mining}},
  author={Saint-Pierre, Cecilia and Cifuentes, Francisco and Bustos-Jim{\'e}nez, Javier},
  booktitle={2014 IEEE conference on communications and network security},
  pages={520--521},
  year={2014},
  organization={IEEE}
}

@inproceedings{dai2025explainableidsiot,
  title={{A Survey of Explainable Intrusion Detection Systems in IoT Networks}},
  author={Dai, Jingze and Huang, Jiaqi and Jiang, Yili and Gyawali, Sohan and Zhong, Fangtian},
  booktitle={International Symposium on Intelligent Computing and Networking},
  pages={420--443},
  year={2025},
  organization={Springer}
}

@article{zebin2022idsxai,
  title={{An explainable AI-based intrusion detection system for DNS over HTTPS (DoH) attacks}},
  author={Zebin, Tahmina and Rezvy, Shahadate and Luo, Yuan},
  journal={IEEE Transactions on Information Forensics and Security},
  volume={17},
  pages={2339--2349},
  year={2022},
  publisher={IEEE}
}

@book{aalst2016pm,
    title = {{Process Mining: Data Science in Action}},
    author    = {van der Aalst, W. M. P},
    year      = {2016},
    edition = {2},
    publisher = {Springer},
    address   = {Berlin, Heidelberg},
    doi = {10.1007/978-3-662-49851-4}}

@article{aldweesh2020deep,
  title={{Deep learning approaches for anomaly-based intrusion detection systems: A survey, taxonomy, and open issues}},
  author={Aldweesh, Arwa and Derhab, Abdelouahid and Emam, Ahmed Z},
  journal={Knowledge-Based Systems},
  volume={189},
  pages={105124},
  year={2020},
  publisher={Elsevier}
}

@misc{vitale2025networktrafficanalysisprocess,
      title={{Network Traffic Analysis with Process Mining: The UPSIDE Case Study}}, 
      author={Francesco Vitale and Paolo Palmiero and Massimiliano Rak and Nicola Mazzocca},
      year={2025},
      eprint={2512.23718},
      archivePrefix={arXiv},
      primaryClass={cs.LG},
      url={https://arxiv.org/abs/2512.23718}, 
}

@article{hariharan2023xai,
  title={{XAI for intrusion detection system: comparing explanations based on global and local scope}},
  author={Hariharan, Swetha and Rejimol Robinson, RR and Prasad, Rendhir R and Thomas, Ciza and Balakrishnan, N},
  journal={Journal of Computer Virology and Hacking Techniques},
  volume={19},
  number={2},
  pages={217--239},
  year={2023},
  publisher={Springer}
}

@article{hozouri2025comprehensive,
  title={{A comprehensive survey on intrusion detection systems with advances in machine learning, deep learning and emerging cybersecurity challenges}},
  author={Hozouri, Ali and Mirzaei, Abbas and Effatparvar, Mehdi},
  journal={Discover Artificial Intelligence},
  volume={5},
  number={1},
  pages={314},
  year={2025},
  publisher={Springer}
}

@article{nandanwar2024dlids,
  title   = {{Deep learning enabled intrusion detection system for Industrial IIoT environment}},
  author  = {Nandanwar, H. and Katarya, R.},
  journal = {Expert Systems with Applications},
  volume  = {249},
  pages   = {123808},
  year    = {2024},
  doi     = {10.1016/j.eswa.2024.123808}
}

@article{elhouda2022idstrustworthiness,
  title   = {{“Why Should I Trust Your IDS?”: An Explainable Deep Learning Framework for Intrusion Detection Systems in Internet of Things Networks}},
  author  = {Abou El Houda, Z. A. E. and Brik, B. and Khoukhi, L.},
  journal = {IEEE Open Journal of the Communications Society},
  volume  = {3},
  pages   = {1164--1176},
  year    = {2022},
  doi     = {10.1109/OJCOMS.2022.3188750}
}

@ARTICLE{nascita2024ntaxaisurvey,
  author={Nascita, Alfredo and Aceto, Giuseppe and Ciuonzo, Domenico and Montieri, Antonio and Persico, Valerio and Pescapé, Antonio},
  journal={IEEE Communications Surveys \& Tutorials}, 
  title={{A Survey on Explainable Artificial Intelligence for Internet Traffic Classification and Prediction, and Intrusion Detection}}, 
  year={2025},
  volume={27},
  number={5},
  pages={3165-3198},
  doi={10.1109/COMST.2024.3504955}}

@inproceedings{Lashkari2017253,
  title={Characterization of tor traffic using time based features},
  author={Lashkari, Arash Habibi and Gil, Gerard Draper and Mamun, Mohammad Saiful Islam and Ghorbani, Ali A},
  booktitle={International conference on information systems security and privacy},
  volume={2},
  pages={253--262},
  year={2017}
}

@inproceedings{leemans2014discovering,
  title={Discovering block-structured process models from incomplete event logs},
  author={Leemans, Sander JJ and Fahland, Dirk and van der Aalst, Wil MP},
  booktitle={International conference on applications and theory of petri nets and concurrency},
  pages={91--110},
  year={2014},
  organization={Springer}
}

@inproceedings{busch2024xsemad,
  title={xsemad: Explainable semantic anomaly detection in event logs using sequence-to-sequence models},
  author={Busch, Kiran and Kampik, Timotheus and Leopold, Henrik},
  booktitle={International Conference on Business Process Management},
  pages={309--327},
  year={2024},
  organization={Springer}
}

@ARTICLE{vitale2025pmdt,
  author={Vitale, Francesco and Guarino, Simone and Flammini, Francesco and Faramondi, Luca and Mazzocca, Nicola and Setola, Roberto},
  journal={IEEE Transactions on Industrial Informatics}, 
  title={Process Mining for Digital Twin Development of Industrial Cyber-Physical Systems}, 
  year={2025},
  volume={21},
  number={1},
  pages={866-875}}

@article{shafi_ntlflowlyzer_2025,
	title = {{{NTLFlowLyzer}: {Towards} generating an intrusion detection dataset and intruders behavior profiling through network and transport layers traffic analysis and pattern extraction}},
	volume = {148},
	issn = {01674048},
	language = {en},
	urldate = {2025-11-19},
	journal = {Computers \& Security},
	author = {Shafi, MohammadMoein and Lashkari, Arash Habibi and Roudsari, Arousha Haghighian},
	month = jan,
	year = {2025},
	pages = {104160},
}

@conference{icissp25,
author={Marta Catillo and Antonio Pecchia and Umberto Villano},
title={{USB-IDS-TC: A Flow-Based Intrusion Detection Dataset of DoS Attacks in Different Network Scenarios}},
booktitle={Proceedings of the 11th International Conference on Information Systems Security and Privacy - Volume 1: ICISSP},
year={2025},
pages={302-309},
publisher={SciTePress},
organization={INSTICC},
doi={10.5220/0013248600003899},
isbn={978-989-758-735-1},
issn={2184-4356},
}

@INPROCEEDINGS{ahmadon2020cpsmqttpbad,
    author={Ahmadon, Mohd Anuaruddin Bin and Yamaguchi, Shingo},
    booktitle={2020 IEEE International Conference on Consumer Electronics (ICCE)}, 
    title={{Process-Based Anomaly Detection and Analysis for Cyber-Physical System with MQTT Protocol}}, 
    year={2020},
    pages={1-6},
    doi={10.1109/ICCE46568.2020.9043165}}

@Article{hornsteiner2024pmopcua,
    AUTHOR = {Hornsteiner, Markus and Empl, Philip and Bunghardt, Timo and Schönig, Stefan},
    TITLE = {{Reading between the Lines: Process Mining on OPC UA Network Data}},
    JOURNAL = {Sensors},
    VOLUME = {24},
    YEAR = {2024},
    NUMBER = {14},
    DOI = {10.3390/s24144497}
}

@article{vitale2025cfad,
title = {{Control-flow anomaly detection by process mining-based feature extraction and dimensionality reduction}},
journal = {Knowledge-Based Systems},
volume = {310},
pages = {112970},
year = {2025},
issn = {0950-7051},
author = {Francesco Vitale and Marco Pegoraro and Wil M.P. {van der Aalst} and Nicola Mazzocca}
}

@article{paolini2025one,
  title={{One-class Anomaly Detection for Industrial Applications: A Comparative Survey and Experimental Study}},
  author={Paolini, Davide and Dini, Pierpaolo and Soldaini, Ettore and Saponara, Sergio},
  journal={Computers},
  volume={14},
  number={7},
  pages={281},
  year={2025},
  publisher={MDPI}
}

@article{malach2025cybershapley,
  title={{CyberShapley: Explanation, prioritization, and triage of cybersecurity alerts using informative graph representation}},
  author={Malach, Alon and Wudali, Prasanna N and Momiyama, Satoru and Furukawa, Jun and Araki, Toshinori and Elovici, Yuval and Shabtai, Asaf},
  journal={Computers \& Security},
  volume={150},
  pages={104270},
  year={2025},
  publisher={Elsevier}
}

@inproceedings{himmelhuber2022detection,
  title={{Detection, explanation and filtering of cyber attacks combining symbolic and sub-symbolic methods}},
  author={Himmelhuber, Anna and Dold, Dominik and Grimm, Stephan and Zillner, Sonia and Runkler, Thomas},
  booktitle={2022 IEEE symposium series on computational intelligence (SSCI)},
  pages={381--388},
  year={2022},
  organization={IEEE}
}

@inproceedings{khalaf2025real,
  title={{Real-Time Detection of Multi-Stage Cyber Attacks in Industrial IoT Networks Using Graph Attention Networks and Temporal LSTM Fusion}},
  author={Khalaf, Qasim M and Al-Attar, Bourair and Pokale, Navnath B and Mohammed, Ali Kadhum and Aljanabi, Yaser Issam Hamodi and Fadhil, Ruaa and Abd Alrazaq, Hayder and Divekar, Neha and Sekhar, Ravi},
  booktitle={2025 3rd International Conference on Cyber Resilience (ICCR)},
  pages={1--8},
  year={2025},
  organization={IEEE}
}

\end{document}